\documentclass{article}

\usepackage{arxiv}

\usepackage[utf8]{inputenc} 
\usepackage[T1]{fontenc}    
\usepackage{hyperref}       
\usepackage{url}            
\usepackage{booktabs}       
\usepackage{amsfonts}       
\usepackage{nicefrac}       
\usepackage{microtype}      
\usepackage{lipsum}

\usepackage{float}
\usepackage{graphicx}
\usepackage{svg}
\usepackage{amsmath}
\usepackage{xcolor,colortbl}
\usepackage{wrapfig}
\usepackage{sidecap}

\newcommand{\circled}[1]{\raisebox{.5pt}{\textcircled{\raisebox{-.9pt} {#1}}}}

\newcommand{\SPOT}{\textsc{spot}}
\newcommand{\tech}[1]{{\small{\textsf{#1}}}}

\title{SPOT: Open Source framework for scientific data repository and interactive visualization}

\author{
  Faruk~Diblen\thanks{Corresponding author} \\
  Netherlands eScience Center \\ 
  Science Park 140 1098 XG Amsterdam \\ 
  The Netherlands \\
  \texttt{f.diblen@esciencecenter.nl} \\
   \And
  Jisk ~Attema \\
  Netherlands eScience Center \\ 
  Science Park 140 1098 XG Amsterdam \\ 
  The Netherlands \\
  \texttt{j.attema@esciencecenter.nl} \\
  \AND
  Rena Bakhshi \\
  Netherlands eScience Center \\ 
  Science Park 140 1098 XG Amsterdam \\ 
  The Netherlands \\
  \texttt{} \\
  \And
  Sascha Caron \\
  Institute for Mathematics, \\
  Astro- and Particle Physics IMAPP, \\
  Radboud Universiteit, \\
  Nijmegen, The Netherlands
  \texttt{email} \\
  \And
  Luc Hendriks \\
  Institute for Mathematics, \\
  Astro- and Particle Physics IMAPP, \\
  Radboud Universiteit, \\
  Nijmegen, The Netherlands
  \texttt{email} \\
  \And
  Bob Stienen \\
  Institute for Mathematics, \\
  Astro- and Particle Physics IMAPP, \\
  Radboud Universiteit, \\
  Nijmegen, The Netherlands
  \texttt{email} \\  
}

\begin{document}
\maketitle

\begin{abstract}
\SPOT{} is an open source and free visual data analytics tool for multi-dimensional data-sets. Its web-based interface allows a quick analysis of complex data interactively. The operations on data such as aggregation and filtering are implemented. The generated charts are responsive and OpenGL supported.
It follows FAIR principles to allow reuse and comparison of the published data-sets. The software also support PostgreSQL database for scalability.
\end{abstract}


\section{Motivation and significance}
\label{sec:motivation}

\begin{SCfigure}
  \centering
           \caption{A typical scientific results from the field of High-Energy-Physics \cite{Aad:2015baa}. A model with 19 parameters  (Supersymmetry) has been tested again the data to determine if parameter sets are excluded by the data or not. Here the fraction of model parameter sets (as colour code) are plotted for two observables predicted by the model as $x$- and $y$-axis. This should show how difficult it is to visualize a 19-dimensional model on a paper and how much information is then lost.}
           \includegraphics[width=0.5\textwidth]{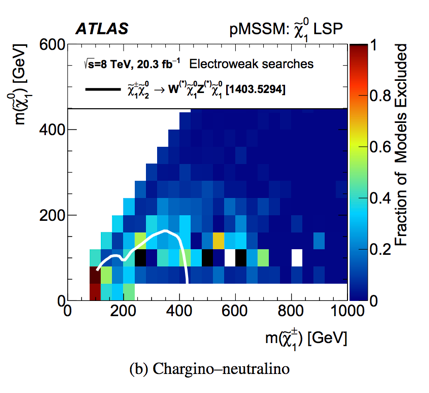}
      \label{fig:example}
  \end{SCfigure}

Most scientific fields produce theoretical or experimental data which is not necessarily the result of a measurement, but also of simulations or evaluations of theoretical models. This data is intrinsically complex consisting of  multiple parameters or multiple observables, and thus, data-sets can be regarded as point clouds in a high-dimensional space. Often, due to the restrictions imposed by the use of paper for data visualization, e.g. a figure in a journal, the status quo is still to publish the data in a two (or three) dimensional format. A typical example is Figure~\ref{fig:example}.

\begin{figure}[!b]
    \centering
      \includegraphics[trim={0 2cm 0 2cm}, width=0.4\textwidth]{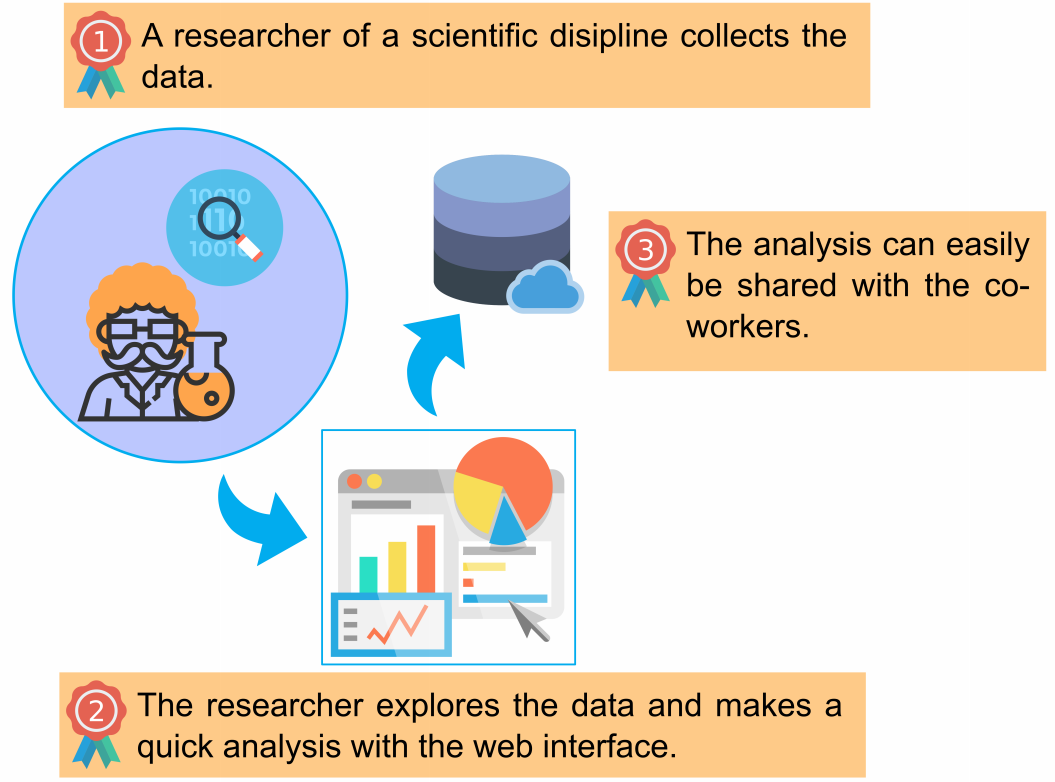}
      \includegraphics[trim={0 2cm 0 2cm}, width=0.4\textwidth]{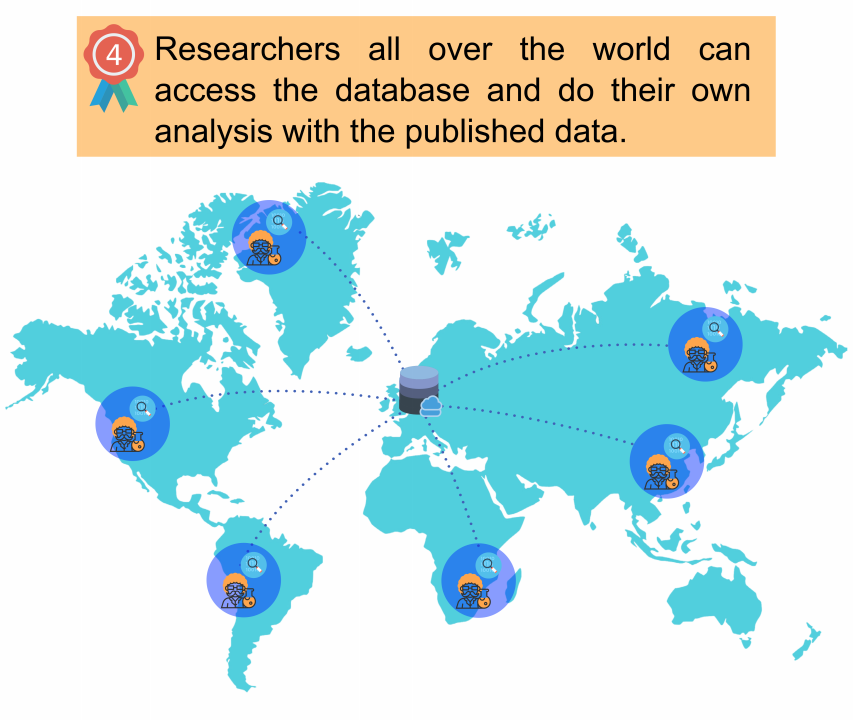}
      \caption{Workflow of \SPOT{}.}
      \label{fig:workflow}
  \end{figure}

However, such a two dimensional (2D) representation obscures most of the correlations within the solution space. 
In order to encourage the publication of high-level data in the complete high-dimensional space, without restrictions, data visualization can be done via web-based tools which allow for, e.g. an automatic generation of multiple relevant histograms. The aim of \SPOT{}~\cite{jisk_attema_2017_1003346} is to provide a flexible data visualization framework to visualize such data. \SPOT{}, which is typically coupled to a database holding the data-sets is a tool to promote the use of \emph{open research data} and \emph{open science}~\cite{machado2015open}. It follows \emph{FAIR}~\cite{wilkinson2016fair} principles to allow reuse and comparison of the published data-sets (see Fig.~\ref{fig:workflow}). This paper briefly introduces \SPOT{}. The source code and the documentation is available at \url{https://github.com/NLeSC/spot}.

\section{Software description}
\label{sec:description}
\SPOT{} provides users with an interactive data exploration environment for high-dimensional data-sets.
The focus is on scientific use, with the aim of facilitating open science, data sharing and reuse (see Figure~\ref{fig:workflow}). It is ideal for numerical data, but categorical (labeled) and temporal data is supported.

Built on a number of concepts from the field of information visualization, it allows a user to create multiple coordinated views called \emph{charts}, showing the data from different perspectives. All charts allow direct manipulation (i.e., selecting and zooming) of the data, and provide visual clues or \emph{animations} when data changes due to user interaction.

\subsection{Software Architecture}
\label{sec:architecture}
The software consists of three components: a \emph{framework}, a \emph{frontend}, and a \emph{server}. A brief description for each of these components follows:

\paragraph{\SPOT{}-framework}
The framework provides classes for data-sets, data views, partitions, aggregation and filtering.
A \emph{data-set} consists of a number of items (or rows), and each item has a set of \emph{facets}.
Facets can be used to partition the data, or they can be aggregated (\textsf{counted}). For numerical data more complex operations are possible, namely, \textsf{summation}, \textsf{averaging}, \textsf{extremes}, \textsf{standard deviation}.
One or more facets make up a \emph{filter}, and all filters combined together form the \emph{data view}.
The user interacts with the framework by setting ranges or selections for the filters, and by adding or removing filters from the data view.
After filtering, the partitioned and aggregated data is available as a simple array, which can be plotted or further processed.
All filters in a data view are linked, and a change in one filter triggers an update of the whole data view.

\paragraph{Frontend}
The frontend is a web-based application with separate pages where a user can upload and define data-sets, a dashboard page that provides the main interaction, and a page where analyses can be downloaded or shared.
The frontend is built on a number of open-source \tech{JavaScript} packages, such as \tech{Chart.js}, \tech{Vis.js}, and \tech{Sigma.js} for animated plots, and \tech{Ampersand.js} and \tech{Material Design Lite} for the interface. 
The filtering is implemented using asynchronous functions, so updates can start as soon as the first results become available. The visualization libraries are \tech{HTML Canvas} based, and offer \tech{OpenGL} accelerated rendering where available.

\paragraph{Server}
The server processes requests for data and applies the necessary filtering and aggregation. When data becomes available, it is pushed to the client which can then update the charts.

We currently have two different implementations for the server component. The first one, which is included in \SPOT-framework, is based on \tech{Crossfilter.js} and runs in the user's web browser without requiring any further resources or even internet access. The second implementation (\SPOT-server), provides a bridge to an external \tech{PostgreSQL} database for scalability. Database queries are run in parallel, and make use of indices for extra performance. Connections to other datastores, like \tech{MySQL} or \tech{MongoDB}, can easily be achieved by extending the server component.

\subsection{Software Functionality}
\label{sec:functionality}

\paragraph{Data import and database connection}
There are two options to import a new data-set.
In the first option, users can upload data available on their own system.
The software supports most common data formats \tech{CSV} and \tech{JSON}, to make data import process easy for different scientific domains.
After the import, the data is checked to automatically detect data types, such as integers and strings. Users can then fix auto detection issues. In the second option, the data is imported from \SPOT-server. The meta data for a data-set (e.g. the name and description) can easily be set in a configuration file stored on server-side.

\paragraph{Dashboard}
\label{sec:charts}
\SPOT{} has eight ready-to-use chart types, namely, horizontal and vertical histograms, line chart, pie chart, bubble chart, 3-d scatter chart, radar chart and network chart. 
Charts are added to the dashboard by clicking the chart icon.
The chart's filter requires one or more facets to partition over, and can take up to 4 facets to aggregate.
Charts show their configuration pane by default.
A visual feature of the chart can be linked to a specific facet by dragging a facet from the top of the screen and dropping it on a slot in the configuration pane.
A Partition or Aggregation can be further configured by clicking on its name on the configuration pane.


\paragraph{Download and share}
The dashboard generated by the user can be saved as a single file, a session file, in \tech{JSON} format. This file contains aggregated data and settings of the dashboard such as existing charts, existing filters. The session file then can be used to restore the analysis. In addition, the session file can be uploaded to a cloud storage and a link to the session file can be shared.

\section{Illustrative Examples}
\label{sec:example}

We show the applicability and the features of the \SPOT{} software on two examples of data-set: (1) the Titanic data-set \cite{titanic-data}, a well-known data-set in data science, and (2) a high-dimensional data-set containing models for dark matter (see \cite{achterberg2017implications,hendriks2015description} for more information).

\begin{figure}[!h]
  \centering
  \includegraphics[width=0.8\textwidth]{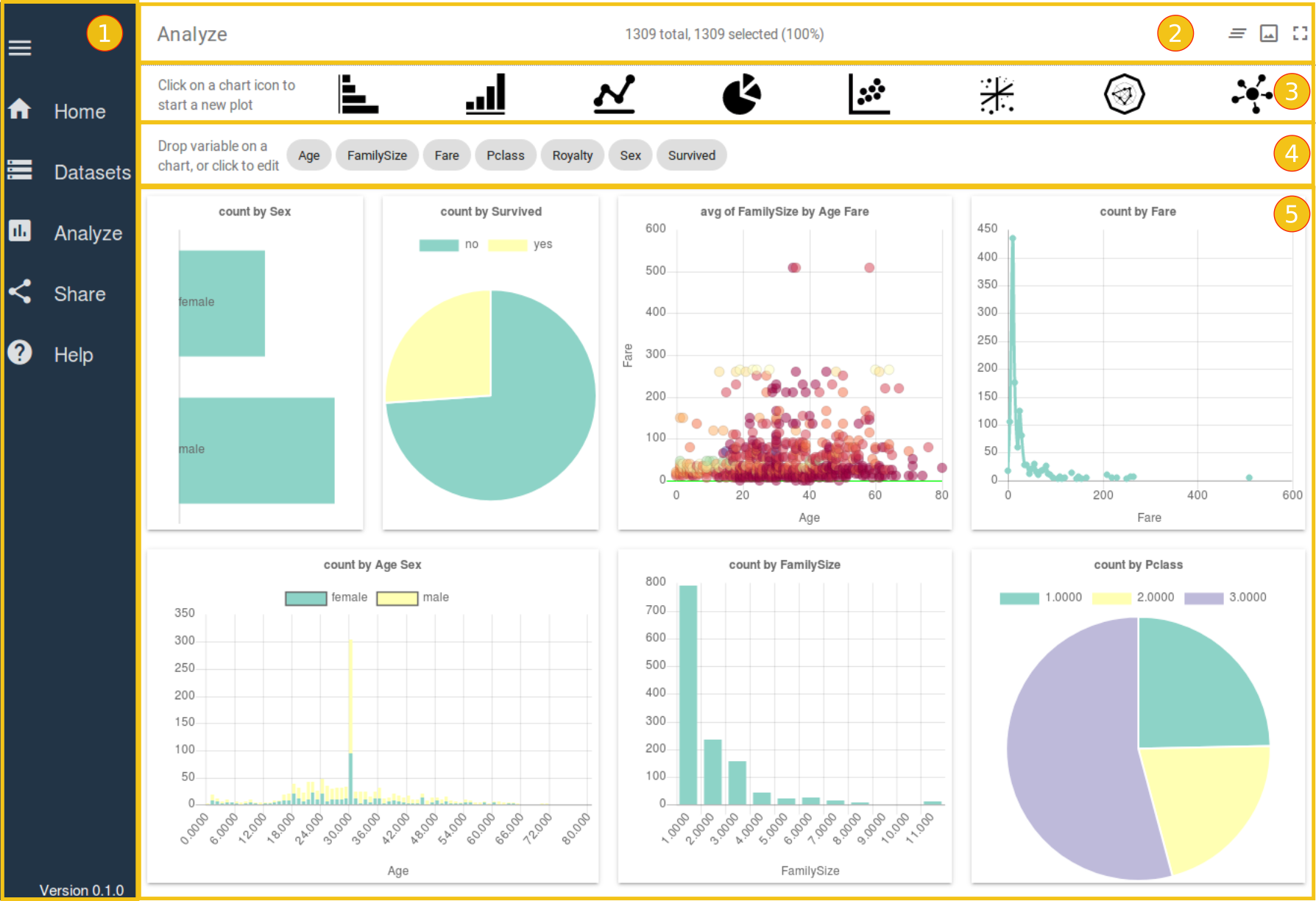}
    \caption{Web-interface of \SPOT{}.
    \circled{1}  Main menu: to navigate between different pages. Users can upload data (Datasets page). analyze it (Analyze page), share selection and charts as a \emph{session} (Share page), and read help documentation (Help page); \circled{2} Information bar:  provides basic statistics about selected data; \circled{3} Chart bar: in order to add available charts to the dashboard; \circled{4} Variable bar: 
      lists available variables within the data-set; \circled{5} The dashboard.}
    \label{fig:titanic}
\end{figure}

\subsection{Titanic data-set}

Visualisation plots of the Titanic data-set are provided in Figure \ref{fig:titanic}; the data can be viewed interactively on the website \url{http://www.idarksurvey.com} by clicking on Demo. 

The top of the figure shows the chart types, each of which can be selected to make a new chart. Directly below that are the data facets, which can be dragged-and-dropped into the empty charts to create any visualisation of any parameter(s). 


\subsection{Example from High-Energy Physics}

\begin{SCfigure}
  \centering
    \includegraphics[width=0.5\textwidth]{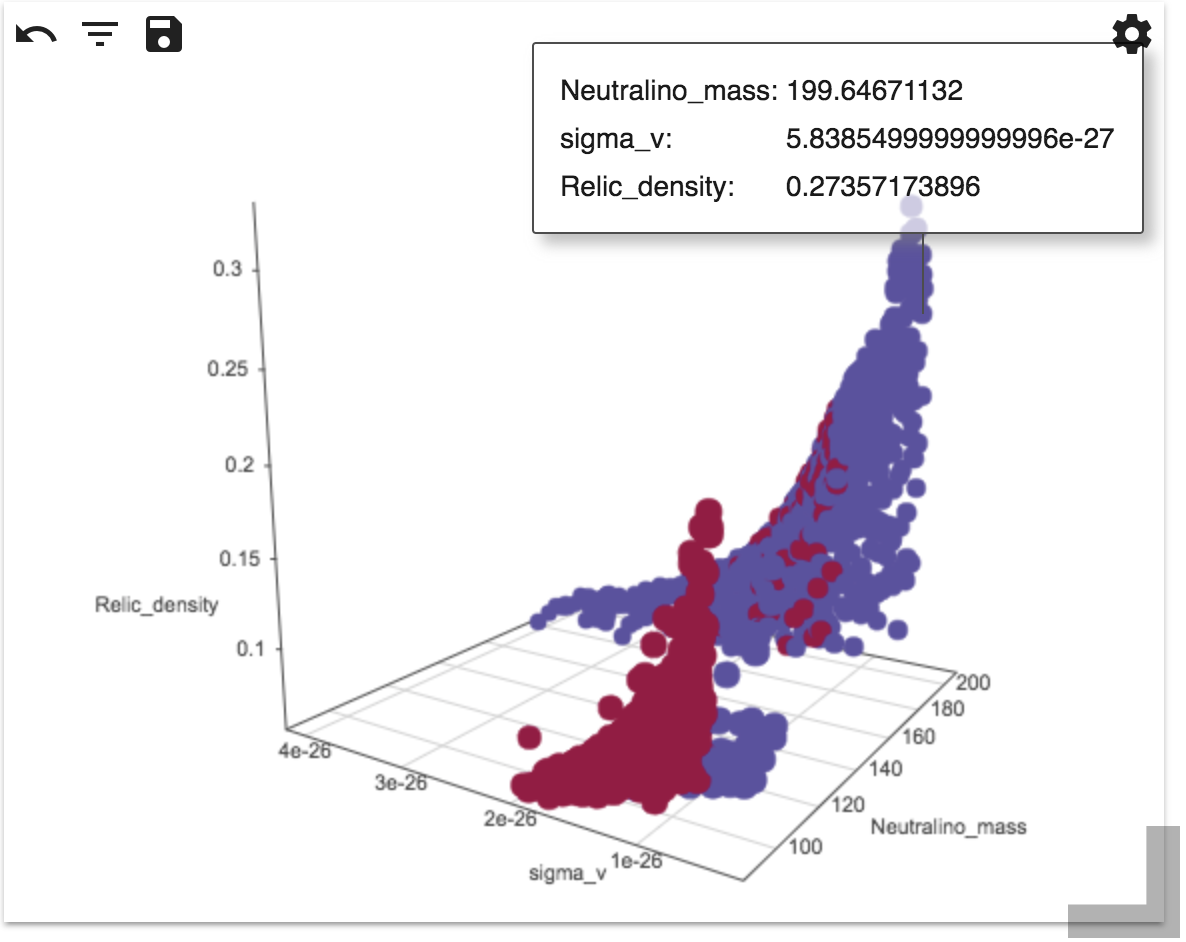}
    \caption{An example 3D visualization of a comparison between two dark matter model data-sets. There are three shared observables. When hovering over a data point, a pop-up tooltip will show the actual values of the model point.}
    \label{fig:dmmodels}
\end{SCfigure}

A real world example where \SPOT{} can help the scientific community is visualizing e.g. models of high-energy physics. The data in this field is typically high-dimensional and even though different models have different theoretical parameters, they share the same observables.
\SPOT{} allows comparison of these observables from different data-sets, providing the user an unprecedented ability to compare  the  model space. In Figure \ref{fig:dmmodels}, two data-sets for  models predicting Dark Matter are compared for three observables: the dark matter mass,  a annihilation probability $\sigma v$ and a cosmic density $\Omega h^2$.

\section{Impact}
\label{sec:impact}

%

The high-dimensional space can be stored in the \SPOT{} database so that the data-set can be published along with the paper. While the paper still contains the most relevant 2D plots, a researcher can plot different variables using \SPOT{} for further research. By intuitively making cuts in for example histograms, researchers can investigate the high-dimensional space in an unprecedented way. In addition, comparison between data-sets on shared observables is likely to lead to new research questions.

Intuitive data visualization is still in its infancy, especially in the field of high energy physics. \SPOT{} aims to be the first tool to provide an intuitive interface for visualizing high-dimensional data-sets and as a place for researcher to store their high-dimensional data.

Thus, there are two categories of applications that are related to \SPOT{}: online sharing services and visualization libraries. The most relevant to \SPOT{} are Microsoft Power BI~\cite{powerbi}, Spotfire~\cite{ahlberg1996spotfire} and Tableau~\cite{heer2008graphical} but these are commercial products. The most popular sharing services include data sharing platform Zenodo~\cite{zenodo}, digital repository for sharing Figshare~\cite{figshare}, and high-energy physics specific platform Hepdata~\cite{maguire2017hepdata}. In comparison with \SPOT{}, however, these services basically provide only storage that allow researchers to upload and publicly store their data and figures, but not compare different visualizations from different articles. On the other hand, visualization frameworks such as Dash~\cite{plotly} is a Python framework that gives users possibility of interactive visualization, but also it requires users to have an advance knowledge to create a desired dashboard.

\section{Conclusions}
\label{sec:conclusion}

The software has been written to provide a free and open state-of-the-art platform for researchers in many scientific domains. It helps researchers to publish, share their data-sets, and collaborate by comparing their data-sets to identify the differences. This makes \SPOT{} a perfect candidate for FAIR data platforms. 

\section*{Acknowledgements}
\label{sec:acks}
This work is supported by the Netherlands eScience Center under the project iDark: The intelligent Dark Matter Survey.

\bibliographystyle{unsrt}
\bibliography{ref}

\end{document}